\documentclass[journal]{IEEEtran}
\usepackage{cite}

\ifCLASSINFOpdf
  \usepackage[pdftex]{graphicx}
\else
\fi
\usepackage{amsmath}

\usepackage{stfloats}

\usepackage{lettrine}
\usepackage{amssymb}
\usepackage{balance}
\usepackage{siunitx}
\usepackage{pgfplots}
\usepackage{csvsimple}
\usepgfplotslibrary{colorbrewer}
\usetikzlibrary{pgfplots.statistics, pgfplots.colorbrewer} 
\pgfplotsset{compat=newest,cycle list/Set1-8}

\usepackage{hyperref}
\usepackage{cleveref}
\newcommand{\mailto}[1]{\href{mailto:#1}{#1}}
\newcommand{\specialcell}[2][c]{%
  \begin{tabular}[#1]{@{}c@{}}#2\end{tabular}}
\DeclareSIUnit{\bit}{bit}
\DeclareSIUnit{\byte}{B}

\begin{document}
%
\title{Versatile Configuration and Control Framework for Real Time Data Acquisition Systems}

\author{N.~Karcher,~R.~Gebauer,~R.~Bauknecht,~R.~Illichmann,~O.~Sander%
\thanks{Manuscript received XXX; revised \today.}%
\thanks{N. Karcher, R. Gebauer, R. Bauknecht, R. Illichmann, O. Sander from Institute for Data Processing and Electronics (IPE), Karlsruhe Institute of Technology (KIT), Karlsruhe, Germany, Email: \mailto{karcher@kit.edu}}%
}



%


\maketitle

\begin{abstract}
Modern physics experiments often utilize FPGA-based systems for real-time data acquisition. Integrated analog electronics demand for complex calibration routines. Furthermore, versatile configuration and control of the whole system is a key requirement. Beside low-level register interface to the FPGA, also access to I$^2$C and SPI buses is often needed to configure the complete system.
Calibration through an FPGA is inflexible and yields a complex hardware implementation. On the contrary, calibration through a remote system is possible but considerably slower due to repetitive network accesses. By using SoC-FPGA solutions with a microprocessor, more sophisticated configuration and calibration solutions, as well as standard remote access protocols, can be efficiently integrated in software. 

Based on Xilinx Zynq US+ SoC-FPGAs, we implemented a versatile control framework. This software framework offers a convenient access to the hardware and a flexible abstraction via remote-procedure calls (RPCs). Based on the open source RPC library gRPC, functionality with low-latent control flow, complex algorithms, data conversions and processing, as well as configuration via external buses can be provided to a client via Ethernet. Furthermore, client interfaces for various programming languages can be generated automatically which eases collaboration among different working groups and integration into existing software. This contribution presents the framework as well as benchmarks regarding latency and data throughput.
\end{abstract}


%
\IEEEpeerreviewmaketitle


\thispagestyle{empty} 
\pagestyle{empty} 

\section{Introduction}
\lettrine[lines=2]{F}{IELD} Programmable Gate Arrays (FPGAs) are nowadays commonly utilized in real-time data acquisition systems of small, medium, and large scale physics experiments. Thereby, the role of FPGAs has expanded to a wide variety of tasks, including data moving, online processing, trigger systems, and more. Configuration and calibration of these systems has become a complex task on its own. Additional components surrounding the FPGA, such as clock chips, optical transceivers, analog-to-digital or digital-to-analog converters, further alleviate the configuration and calibration complexity. These tasks could be implemented inside the FPGA. However, this would require a lot of implementation effort, take a considerable amount of resources, and only provide limited flexibility. 

Modern heterogeneous FPGAs integrate traditional CPU cores with an FPGA fabric into a so-called multi-processor system-on-chip (MPSoC), thereby allowing for a customized partitioning of the given workload into software and hardware. Primarily non-time-critical control-flow driven configuration and calibration tasks can be handled in software. Quicker implementation time and features such as advanced arithmetic support with hardware division or floating-point calculations come in handy for these tasks. Also, remote monitoring and control access to DAQ systems is mandatory for larger experiments that can contain dozens of such platforms and measure over a longer time period.

We designed and implemented a versatile control framework for DAQ systems based on heterogeneous MPSoCs. Our development primarily targets the application domains concerning the readout of superconducting sensors and quantum bits. Both fields need highly customized FPGA processing due to performance and latency requirements. At the same time, both applications utilize complex calibration procedures which are better suited for software implementation. Our software framework offers convenient access to the resources of the system via remote-procedure calls (RPC). These offer a flexible abstraction as they can either execute a low-level access or complete calibration sequences. Utilizing the open-source RPC library \emph{gRPC} the client can connect to the provided functionality. It also allows to automatically generate a client interface for languages like Python, Go, C/C++, Rust, C\# and many more. 

The main component of the presented implemented software framework is the ServiceHub daemon. The ServiceHub, including its plugins, can be used to control and configure the platform, as well as acquire data with light to medium data rates. The key benefits of this approach based on a SoC-FPGA combined with the gRPC system is a modular reliable control solution with a flexible abstraction of functionality. It offers easy integration in existing client applications and encapsulates fast control flow and math on the device itself. Furthermore, due to gRPC features, it allows users to implement a client in their preferred language.

\section{Fundamentals}
\label{sec:fundamentals}

\subsection{Heterogeneous SoC-FPGA}

The two largest FPGA manufactures, Xilinx and Intel, offer System-on-Chip devices that incorporate large Field-Programmable-Gate-Arrays (FPGA) for custom logic implementation and CPU cores for custom software. For the targeted DAQ systems, the Xilinx Zynq UltraScale+\cite{XilinxUSP} SoC-FPGA family is used. It consists of two partitions: the Processing System (PS) and the Programmable Logic (PL). The PS integrates a four core CPU (ARM Cortex A53), a two core RPU (ARM Cortex R5), GPU and various interface controllers such as Ethernet MACs, I$^2$C and SPI bus masters, and a DDR4 controller. The hardware units are interconnected via an AXI bus system. The PS side can run a fully featured operating system and is supported by the Linux Kernel.
The PL offers an FPGA with additional hard ip cores, dependent on the device type. The PS connects to the PL with three AXI master, nine slave interfaces, clocks and other connections like interrupt lines. AXI compatible devices in the PL can be configured to share the physical address space with the PS. This allows to access devices through memory mapping, hence to access it with Linux kernel platform drivers or through userspace drivers.

\subsection{Remote Procedure Calls}
\label{sec:rpc}
Software following the procedural programming paradigm contains the concept of procedure calls, which encapsulate a given piece of functionality. The encapsulated code can be parameterized with function parameters and eventually returns a result to the calling context. Remote Procedure Calls (RPC) separate the calling context and the executing context in to different processes or even computer systems, allowing an inter-process communication (IPC). Local procedure calls are cheap, whereas remote procedure calls are not\cite{rfc707} -- Arguments and return values for remote procedure calls need to be serialized and made platform independent to bridge between two different hosts. Hereby marshalling is used to make data types platform independent. Network access adds an overhead and therefore additional call latency.
In contrast, two different processes can be seamlessly coupled by RPC which eases implementation of a distributed system. \emph{gRPC}\cite{grpcgit} is an RPC library. It is based on the \emph{protobuf} library and was initially developed by Google. Both libraries together offer serialization, code generation and additional security layers. GRPC is based on HTTP/2.0 over TCP/IP and transfers the data in a binary protobuf format. The calls over TCP/IP can be affected by retransmission that could lead to undeterministic latency. This counts for any protocol on standard Ethernet, without any additional hardware\cite{7005074}. Software with tighter timing constraints should be run directly on the device. It proves to be advantageous that the flow control offers a reliable data transfer as well as connection monitoring. Deadlines for an RPC call can be defined before a call in order to detect a connection timeout.

The two basic principles of gRPC for describing the interface definition are RPC functions and \emph{protobuf messages}. The latter defines the structure of call arguments and return values. Both are defined by the developer inside a protocol definition file. The latter is used to generate both interface sides, the \emph{stub}-classes for calling and \emph{base}-classes (or \emph{Servicer}) for called context. Hereby the generator is able create files for various programming languages, making the server agnostic to the client language. A main advantage between gRPC and other RPC solutions is the wide support of target languages, from C up to JavaScript. GRPC provides bidirectional streaming and binary data format and has a very active development group, as well as a rich documentation.

\subsection{Target System-Architectures}
The control software presented in this paper was developed for readout system for metallic magnetic calorimeters and superconducting quantum bits (qubits).
In the context of experiments with these special superconducting circuits, the data acquisition system consists of three partitions: a frequency conversion, a digital-to-analog conversion, and a data processing system. Each partition needs to be configured and calibrated before, or in predefined intervals during, measurements. For repetitive calibrations, data is processed in order to configure external chips via I$^2$C and SPI; in addition, the FPGA IP Cores are reconfigured to achieve an iterative optimization. This interplay requires fast feedback of the calibration results to configuration values to prevent downtime during measurement, which would cause an increased absolute run time.

\begin{figure}
    \centering
    \includegraphics[width=0.8\columnwidth]{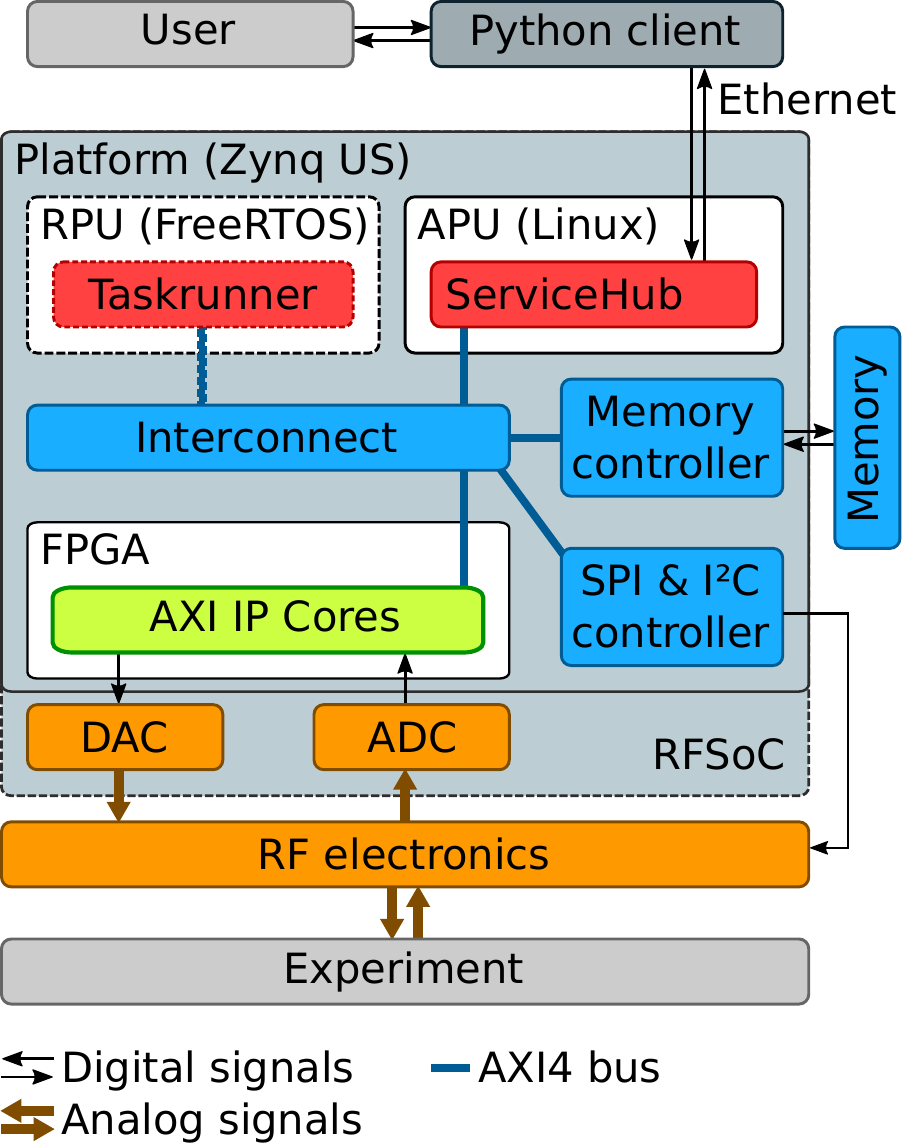}
    \caption{Combined platform diagram of the qubit and micro calorimeter readout system. The qubit readout uses the RFSoC family that integrates the ADC/DAC stage, whereas the calorimeter readout uses an external conversion stage. The qubit readout system also integrates ARM R5 core to render real-time control flows possible through the Taskrunner framework\cite{gebauer2020accelerating}.}
    \label{fig:target-platform-combined}
\end{figure}

\Cref{fig:target-platform-combined} contains a block diagram of the generalized readout system concept for qubits and metallic magnetic calorimeters. Although the physical system components  are similar, the internal firmware and SoC usage is significantly different.
For the readout of metallic magnetic calorimeters, which is one type of superconducting sensors, the platform follows a streaming architecture. It can acquire and process a continuous data stream in real-time, filters, demodulates and triggers to evaluate 8000 Events/s in a raw-data stream of \SI{20}{\giga\byte\per\second}. The real-time processing is exclusively done on the FPGA side of the device on a sample-per-sample basis. The processing system side of the SoC is used for calibration, control, and data transfer. During processing inside the FPGA partition, the stream is reduced to a moderate rate below \SI{80}{\mega\byte\per\second}\cite{8706975} that can be sent to a backend server via a ServiceHub plugin.

For the readout and control of superconducting quantum bits, it is necessary to generate and evaluate microwave pulses. The major difference in firmware is that the acquisition is not continuous but in repetitive control and readout intervals. It also separates the task between the application processor, the real-time processor, and a sequencer on the FPGA to achieve different accuracy levels, down to the nanosecond timescale for pulse generation and acquisition\cite{gebauer2020accelerating,doi:10.1063/5.0011721}. The system is mainly used as measurement equipment to perform basic research on superconducting qubits and determine their characteristics. The gRPC protocol interface of the platform allows easy integration into the python-based measurement framework Qkit\cite{qkit}, that configures the device and gathers the data. Also more complex driver capabilities are implemented inside the ServiceHub plugins, thereby providing a better abstraction, reducing the complexity of client-side drivers and making them easier to maintain.

\subsection{Related Work}

The ServiceHub framework is a middleware between the hardware and the client application. It should be distinguished from pure system monitoring software like IPMI, OpenBMC or u-bmc. These programs run in parallel to the actual operating system and are often executed on co-processors with their own operating system and kernel. The ServiceHub is dependent on a running Linux operating system.

A common control instrument for FPGA systems is the IP-Bus\cite{Larrea_2015} which implements a protocol on top of UDP/IP. The IP-Bus provides pure register resources of an FPGA via network and is characterized by a low round-trip latency of \SI{250}{\micro\second}. However, the protocol offers only limited possibilities for complex calibrations without frequent network access, as this is done with additionally provided software \emph{ControlHub} and \emph{uHAL} which are executed on a remote computer system.

The closest representative to the system presented can be found in the protocols for the control of measurement equipment. Besides the well-known representative LXI, there is the RPC-based VXI-11 protocol. This enables the generation of types, stubs and basic structures in C. However, it uses these features to implement the IEEE 488.2 standard. Therefore a seamless integration in software is not possible, it is usually integrated via VISA. VXI-11 is now considered obsolete and is replaced by the non-RPC based HiSLIP.

Beside the instrument control protocols, the EPICS software can be found in distributed systems of accelerator physics and was also migrated to the Zynq platform. EPICS focuses on slow control. It utilizes process variables (PV) that can represent values in memory or of control or monitor devices\cite{osti_6110347}. The variables can be fetched, changed and monitored with \emph{Channel Access} or the newer \emph{pvAccess} module. \emph{pvAccess} also allows to implement modern service based drivers and extend them with a custom RPC functionality. EPICS IOC was already migrated to Zynq\cite{7543117}. The IOC is usually used in a distributed EPICS network for slow control\cite{osti_6110347} and sharing of process variables, whereas our system focuses on efficient and fast peer-to-peer access and data transfer. 

GRPC based solutions are not commonly used on embedded SoC-FPGAs, but there is research related to hardware acceleration techniques in which gRPC is used to distribute tasks among FPGA platforms\cite{8666151}.

\section{ServiceHub Software}

The ServiceHub forms a configurable, modular plugin loader offering a standardized interface for gRPC function registration, logging, health and plugin management. A plugin is an entity for controlling a specific device or software class, including FPGA IP cores with AXI interface, external devices attached to the SoC or even other plugins within the ServiceHub.

\begin{figure}
    \centering
    \includegraphics[width=0.95\columnwidth]{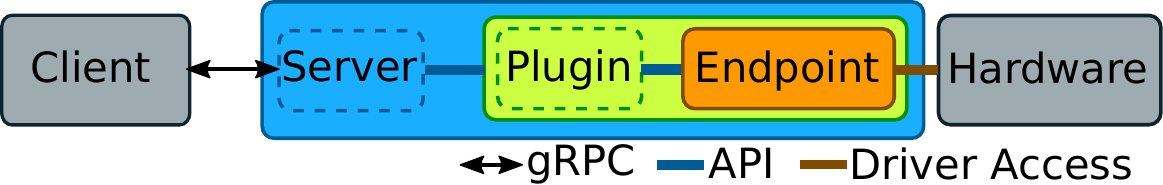}
    \caption{Core principle of operation of the ServiceHub}
    \label{fig:servicehub-idea}
\end{figure}

The idea of the ServiceHub is shown in \cref{fig:servicehub-idea}. The client communicates with a single gRPC server. The functionality of the daemon is kept modular and is encapsulated in plugins. All functionality of a plugin is implemented in a C++ class from which functions for remote configuration are exposed to the client-side. These are defined in gRPC protocol files that are eventually also used to generate stub classes which are inserted in a client application. The plugins are loaded if defined in the main configuration. A plugin may include one or multiple so-called endpoints which implement the drivers to access underlying hardware devices. One endpoint per physical device is instantiated automatically via the device tree of the Linux kernel. A plugin can control other plugins, this allows to coordinate the interplay between different plugins, e.g. to conduct calibrations.

\subsection{Plugin Structure and Mechanism}

A ServiceHub plugin is a class that inherits from two other classes. The functional part inherits from a service-class, which is provided by gRPC via code generation. This generated code originates from the protocol file describing the interface between RPC client and server. The second, virtual base class qualifies the class as a ServiceHub plugin. This ensures that every plugin implements the same set of basic functionality, allowing the ServiceHub to handle every plugin the same way. Each plugin class can be instantiated with exported symbols of wrapper functions for constructor and destructor of the plugin. This ensures the plugin can be loaded properly by the server executable. A macro function for eased exporting is provided.

\begin{figure}
    \centering
    \includegraphics[width=0.95\columnwidth]{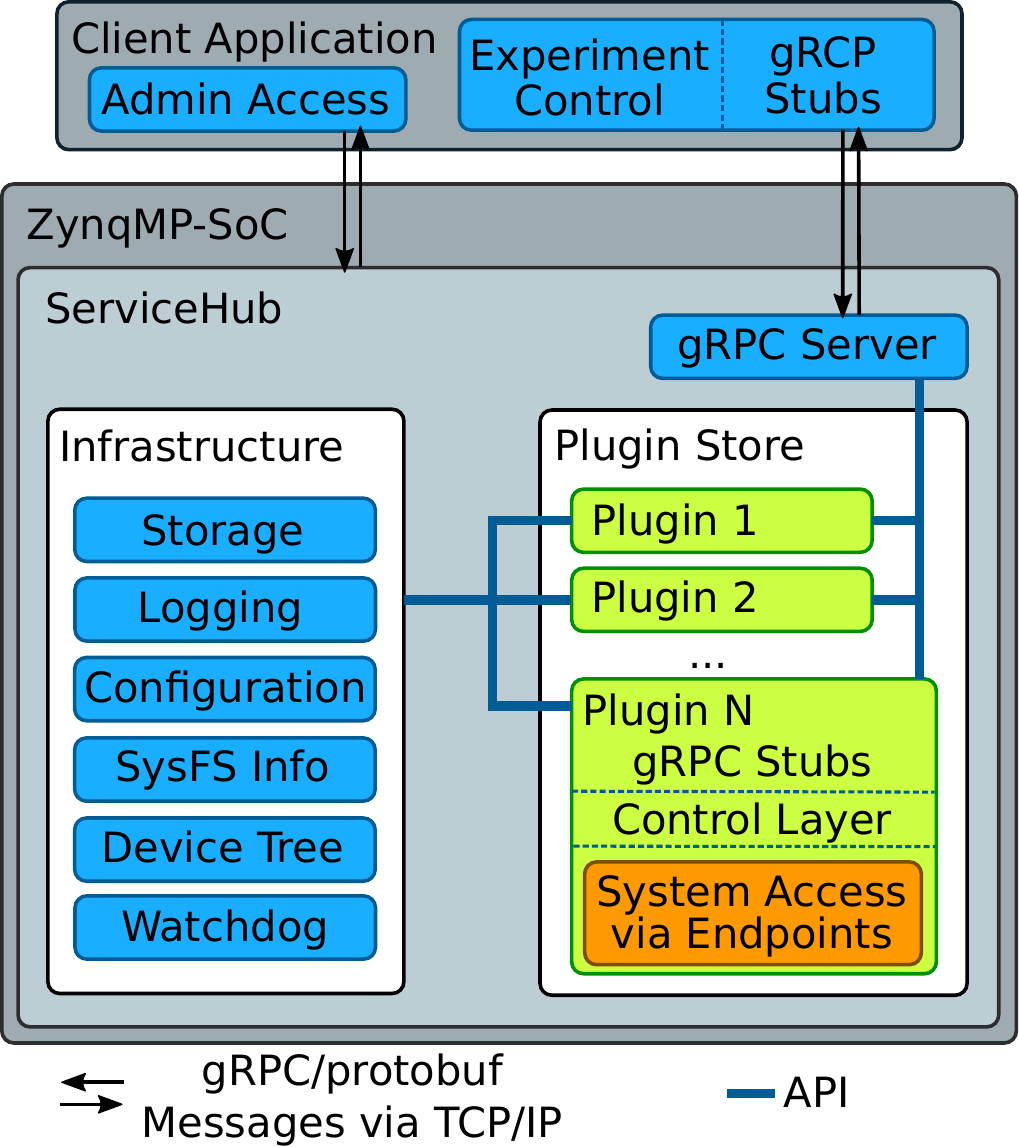}
    \caption{Structure of the ServiceHub host and the controlled plugins. The admin access is allowed by a gRPC service internal to the host.}
    \label{fig:servicehub-structure}
\end{figure}

ServiceHub plugins can be created without full access to the ServiceHub code. A common set of headers forms a template for plugins with a compatible interface. This allows the ServiceHub to handle all plugins the same way while keeping a strict separation between plugin and server code without affecting any functionality of the gRPC protocol. The internal structure of the program is shown in \cref{fig:servicehub-structure}.

The functionality of a plugin can be split up into two layers. The \emph{access layer} provides an interface to the hardware or system resources which are needed to fulfill the plugin task. Hardware resources are usually encapsulated in a so-called \emph{endpoint} type, which is described more closely in \cref{sec:hwinterface}. The plugin task is defined by the code in the \emph{control layer}. Decisions, higher level algorithms and plugin-plugin interactions are usually implemented there. 

When the ServiceHub server is started, it first reads its configuration file using the \emph{json} library by Niels Lohmann \cite{jsongit}. Next to server configuration values, this file also includes a list of plugin requests and plugin-specific configuration values. To keep flexibility, plugins are loaded at run-time and implement the functionality available to the user, internally as well as externally via gRPC. For each requested plugin, the server tries to locate the plugin library and to load the exported symbols for plugin creation and deletion. This is done via the dynamic linker interface given in the \texttt{dlfcn.h} header. If dynamic linkage is successful, the logger for the plugin is created and the constructor is called with the previously described arguments.

At instantiation, every plugin object is assigned a logger and a configuration. The logger is dedicated to the plugin and based on the spdlog library \cite{spdloggit}. The configuration is a segment of the full configuration file.
Additionally, plugin objects receive a pointer of an interface type which allows limited access to ServiceHub resources such as a list of currently loaded plugins.

After instantiation, the plugin objects are added to a gRPC server instance, which handles all network communication after the startup has finished. While running, the server accepts all RPC requests for the loaded plugins and relays them to the dynamically loaded code.

Plugins can be controlled by other plugins to enhance reusability. To achieve this, a plugin can request access to currently loaded plugins from the ServiceHub. The requested plugins can then be controlled via their public interface.

\subsection{Hardware and Kernel Access}
\label{sec:hwinterface}
The ServiceHub framework offers a set of functions and base classes to enable hardware access. The functions are summarized in so-called endpoint types, the parent classes. 

An endpoint type defines the type of access: For AXI registers the endpoint is instantiated with a memory map (\texttt{PlatformEndpoint}), to control kernel drivers a file IO is used in the endpoint (\texttt{SysfsEndpoint}). New final endpoints can be implemented by deriving the types. Each endpoint type brings along a static factory method which is called in the constructor of the plugin. The connection between the hardware and the instances of the final endpoint is established via the device tree. For each entry in the device tree an instance is created. The device tree can be used to read out the address ranges, and the device tree node can be compared with the sysfs paths to find the path to the Linux device driver.

The access functions are implemented as template functions. Different types can be read or written with \texttt{read\allowbreak<type>(address)}, \texttt{write<>(address, var)}. Furthermore, overloaded functions for bitwise access are offered. For Linux device drivers the node name in the sysfs must be specified instead of an address. The Linux Kernel drivers allow access to external devices (I$^2$C, SPI) or devices that require interrupt support.

\subsection{Reliability}
Experiments that are operated over a long period of time require a reliable monitoring of the driver components and the hardware. The ServiceHub offers functions to enable plugins to report their health status and, if necessary, to register with a software watchdog. If one of the endpoints within a plugin is no longer functional, the endpoints can be reloaded without stopping the gRPC server.

The ServiceHub registers with the hardware watchdog of the Zynq-SoC via the watchdog kernel driver. If the ServiceHub does not respond due to a malfunction, the system will be rebooted.

An optional plugin is provided which offers monitoring of relevant system parameters such as temperature, voltages and boot data integrity via check sums. The plugin features an interface to the monitoring tool monit as well \cite{monitgit}. If activated, this allows to further review the system state regarding resource utilization and executed processes. The plugin is notified by monit if an issue is detected and acts accordingly.

\section{Performance Evaluation}
As middleware, the ServiceHub encapsulates functionality on the device and interacts with the hardware registers with the lowest possible latency. On the other hand, the RPCs are executed via network, where a much higher latency can be expected. For a comprehensive analysis, the access times of the memory-mapped registers are checked first, followed by the network accesses. Subsequently, a throughput measurement is presented. Finally, the slow control connection via the kernel drivers is evaluated. 

\subsection{Measurement Setup}
A Xilinx ZCU102 evaluation board with a Zynq US+ SoC was used for the measurement. The board is connected via \SI{1}{\giga\bit\per\second} Ethernet to a client PC without any additional switches or routers. On the host PC a CentOS 7 with more recent GCC 7.2 devtoolset was used to build and run the necessary libraries and client application. On the target system a Yocto Linux based on the 2018.3 version branch of Xilinx was installed. The ServiceHub was compiled with GCC 7.3. For both the host- and the cross-compiler, the compiler flag for code optimization (\emph{-O2}) was enabled. During execution, the application on host and client is changed to highest priority (\emph{-20}) and \emph{FIFO} scheduling, in order to minimize disturbance due to other processes. A Linux kernel real-time patch set(\emph{PREEMPT\_RT}) was not used.

The hardware of the target system consists of an AXI4 interconnect which is connected to the HPM0 master port of the PS. A generic AXI4Lite slave, which is used for register benchmarking, is connected to the interconnect. There are four slave devices connected to it in total. From ZynqMP to Interconnect a 128 bit wide interface is used. The AXI4Lite DUT slave has a \SI{32}{\bit} interface. The AXI bus system on the PL side is clocked with \SI{125}{\mega\hertz} derived from the IOCLK.

\subsection{Register Access}
Access to AXI register is usually used to configure IP cores, query status or transfer small amounts of data. In the measurements, the access times to \SI{32}{\bit} registers of the AXI4Lite module are examined. AXI4Lite is a simplified version of the AXI4 interface, which does not support certain features like burst read access. This saves resources at the expense of data throughput. During the benchmark, single register accesses are performed without ascending or descending addresses. This avoids burst reads in the AXI4 infrastructure within the PS.

To estimate the additional latency caused by the Linux operating system, the ServiceHub, and the endpoint structure, the measurement is performed in three ways. First, using a bare metal program on the ARM Cortex A53, second the pointer access with the Linux memory map and third inside an endpoint with the endpoint API located on the top level of the software stack.



\begin{table}[h]
    \centering
    \caption{Single \SI{32}{\bit} latencies. Comparison between accesses in bare metal software, with Linux MMAP pointer, with endpoint API and from client via Ethernet.}
    \begin{tabular}{p{1cm}|c|c|c|c}
    & \textbf{Bare Metal} & \textbf{\specialcell{Linux\\MMAP}} & \textbf{Endpoint} & \textbf{Client}  \\ 
    \hline
    Read(R) & \SI{299(4)}{\nano\second} & \SI{326(7)}{\nano\second} & \SI{482(17)}{\nano\second} & \SI{274(16)}{\micro\second}  \\ 
    Write(W) & \SI{210(2)}{\nano\second} & \SI{243(9)}{\nano\second} & \SI{280(99)}{\nano\second} & \SI{262(15)}{\micro\second}  \\ 
    W\&R & & & \SI{721(18)}{\nano\second} & \SI{254(9)}{\micro\second}
    \end{tabular}
    \label{tab:register-latency}
\end{table}

From the measured values in \cref{tab:register-latency} one can recognize that the memory mapping in the Linux operating system produces a light overhead, compared to bare metal. The memory is mapped with the parameter O\_SYNC, and for write operations the intrinsic \emph{\_\_dsb} was used to flush the data. On the endpoint level, the latency is further increased as it includes additional function calls.

\begin{figure}
    \centering
    \resizebox{0.95\columnwidth}{!}{%
    \begin{tikzpicture}
        \begin{axis}[xmode=log,log basis x={2},ymode=log,log basis y={10},
            xlabel={Array size},xtick={1,4,16,64,256,1024,4096,16384,65536,262144},
            ylabel={$t/\mathrm{ns}$},
            legend pos={north west},
        ]
        \addplot [color=blue, only marks,mark=o,]
                 plot [error bars/.cd, y dir = both, y explicit]
                 table[x index=0, y index=2, y error index=3, col sep=comma]{grpc_singlecall32bitRead.csv};
        \addlegendentry{read}
        \addplot [color=blue!50!yellow, only marks,mark=+,]
                 plot [error bars/.cd, y dir = both, y explicit]
                 table[x index=0, y index=2, y error index=3, col sep=comma]{grpc_singlecall32bitWrite.csv};
        \addlegendentry{write}
       \end{axis}
    \end{tikzpicture}%
    }
    \caption{Endpoint \SI{32}{\bit} access latencies depending on array length. Read latency is significantly larger than write latency. The larger error bar for the second write measurement results from spikes in the measurement (compare \cref{fig:endpoint-singlecall-boxplot}).  (N=1000 averages)}
    \label{fig:endpoint-singlecall}
\end{figure}
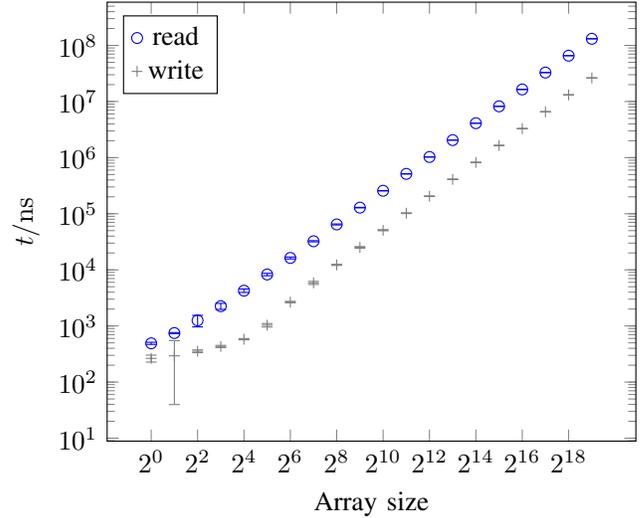

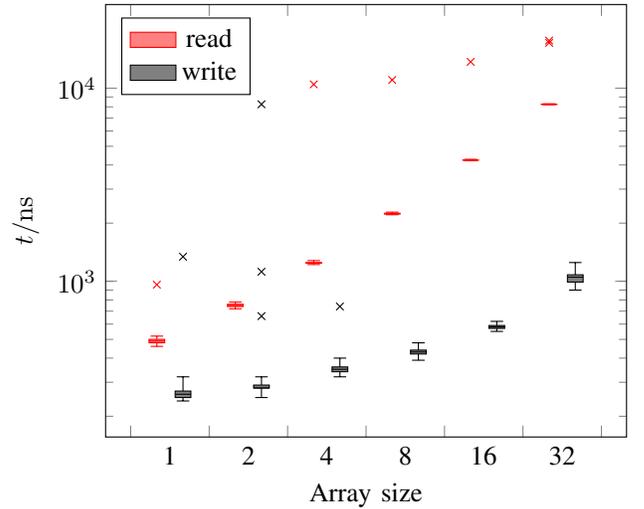
\begin{figure}
    \centering
    \resizebox{0.95\columnwidth}{!}{%
    \begin{tikzpicture}
        \begin{axis}[ymode=log,log basis y={10},
            boxplot/draw direction=y,
            xlabel={Array size},
            xtick={0,1,2,...,50},
            x tick label as interval,
            xticklabels={1,2,4,8,16,32},
            ylabel={$t/\mathrm{ns}$},
            legend pos={north west},
            cycle list={{red},{black}},
            area legend,
            mark = x,
            boxplot={
                  %
                  %
                  draw position={1/3 + floor(\plotnumofactualtype/2) + 1/3*mod(\plotnumofactualtype,2)},
                  %
                  box extend=0.2,
              },
            every axis plot/.append style={fill,fill opacity=0.5},
        ]
          \addplot +[mark=x,boxplot prepared={lower whisker=460.000000,lower quartile=480.000000,median=490.000000, upper quartile=500.000000,upper whisker=520.000000},] coordinates {
(0,960)
};
\addplot +[mark=x,boxplot prepared={lower whisker=240.000000,lower quartile=250.000000,median=260.000000, upper quartile=270.000000,upper whisker=320.000000},] coordinates {
(0,1340)
};
\addplot +[mark=x,boxplot prepared={lower whisker=720.000000,lower quartile=740.000000,median=750.000000, upper quartile=760.000000,upper whisker=780.000000},] coordinates {
};
\addplot +[mark=x,boxplot prepared={lower whisker=250.000000,lower quartile=280.000000,median=280.000000, upper quartile=290.000000,upper whisker=320.000000},] coordinates {
(0,8250)
(0,1120)
(0,660)
};
\addplot +[mark=x,boxplot prepared={lower whisker=1220.000000,lower quartile=1240.000000,median=1240.000000, upper quartile=1250.000000,upper whisker=1280.000000},] coordinates {
(0,10460)
};
\addplot +[mark=x,boxplot prepared={lower whisker=320.000000,lower quartile=340.000000,median=350.000000, upper quartile=360.000000,upper whisker=400.000000},] coordinates {
(0,740)
};
\addplot +[mark=x,boxplot prepared={lower whisker=2210.000000,lower quartile=2230.000000,median=2240.000000, upper quartile=2250.000000,upper whisker=2280.000000},] coordinates {
(0,11020)
};
\addplot +[mark=x,boxplot prepared={lower whisker=390.000000,lower quartile=420.000000,median=430.000000, upper quartile=440.000000,upper whisker=480.000000},] coordinates {
};
\addplot +[mark=x,boxplot prepared={lower whisker=4220.000000,lower quartile=4240.000000,median=4240.000000, upper quartile=4250.000000,upper whisker=4280.000000},] coordinates {
(0,13680)
};
\addplot +[mark=x,boxplot prepared={lower whisker=550.000000,lower quartile=570.000000,median=580.000000, upper quartile=590.000000,upper whisker=620.000000},] coordinates {
};
\addplot +[mark=x,boxplot prepared={lower whisker=8211.000000,lower quartile=8240.000000,median=8250.000000, upper quartile=8261.000000,upper whisker=8300.100000},] coordinates {
(0,17631)
(0,17160)
};
\addlegendentry{read}
\addplot +[mark=x,boxplot prepared={lower whisker=900.000000,lower quartile=990.000000,median=1050.000000, upper quartile=1080.000000,upper whisker=1250.000000},] coordinates {
};
\addlegendentry{write}

       \end{axis}
    \end{tikzpicture}%
    }
    \caption{Endpoint \SI{32}{\bit} access latency boxplot for small transfer sizes. The box represents the quartiles, the whiskers the \SI{99}{\percent}-quantile. Outliers (indicated by the additional markers) occurred for read and write operations and are in the range of \SI{10}{\micro\second}. (N=1000 averages)}
    \label{fig:endpoint-singlecall-boxplot}
\end{figure}

In \cref{fig:endpoint-singlecall} the latency is shown as a function of the transfer size measured using the \SI{32}{\bit} endpoint access. The latency first increases non-linearly due to the overhead in the Endpoint API before entering a linear regime for larger transfers. In  \cref{fig:endpoint-singlecall-boxplot} the latency distribution for small transfer sizes is shown. All latency measurements are strongly affected by outliers in the range of \SI{10}{\micro\second} which are likely caused by other processes within Linux. The standard deviation is considerably improved for bare metal software (see \cref{tab:register-latency}).

\begin{figure}
    \centering
    \resizebox{0.95\columnwidth}{!}{%
    \begin{tikzpicture}
        \begin{axis}[xmode=log,log basis x={2},ymode=log,log basis y={10},
            xlabel={Array size},xtick={1,4,16,64,256,1024,4096,16384,65536,262144},
            ylabel={$t/\mathrm{ns}$},
            legend pos={north west},
        ]
        \addplot [color=blue, only marks,mark=o,]
                 plot [error bars/.cd, y dir = both, y explicit]
                 table[x index=0, y index=9, y error index=10, col sep=comma]{grpc_singlecall32bitRead.csv};
        \addlegendentry{hardware read}
        \addplot [color=blue!50!yellow, only marks,mark=o,]
                 plot [error bars/.cd, y dir = both, y explicit]
                 table[x index=0, y index=9, y error index=10, col sep=comma]{grpc_singlecall32bitWrite.csv};
        \addlegendentry{hardware write}
        \addplot [color=red, only marks,mark=+,]
                 plot [error bars/.cd, y dir = both, y explicit]
                 table[x index=0, y index=9, y error index=10, col sep=comma]{grpc_singlecall32bitReadSW.csv};
        \addlegendentry{software  read}
       \end{axis}
    \end{tikzpicture}%
    }
    \caption{Client access \SI{32}{\bit} access latencies depending on array length. This latency involves network access over gRPC, therefore is significantly higher than in Figure \ref{fig:endpoint-singlecall}. For smaller package sizes the network access dominates the latency. For larger array sizes the endpoint bandwidth gets significant and limits the data throughput thereby the access latency. The single call throughput to hardware is read \SI{81}{\mega\bit\per\second} and write \SI{140}{\mega\bit\per\second}. -- in comparison for a write to a variable in software it is \SI{314}{\mega\bit\per\second}. (N=1000 averages)}
    \label{fig:client-singlecall}
\end{figure}
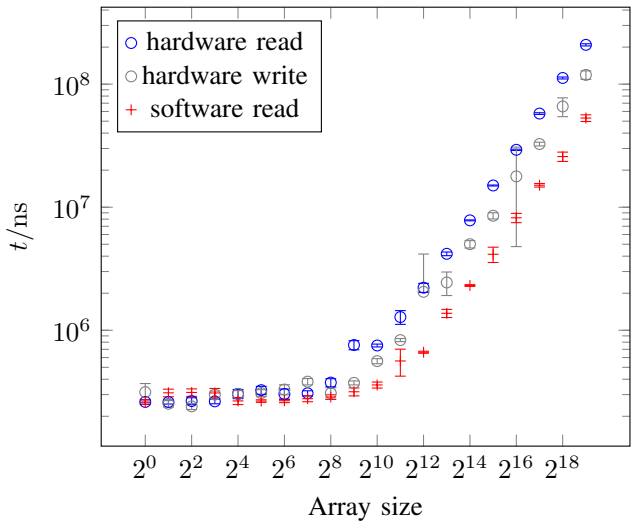

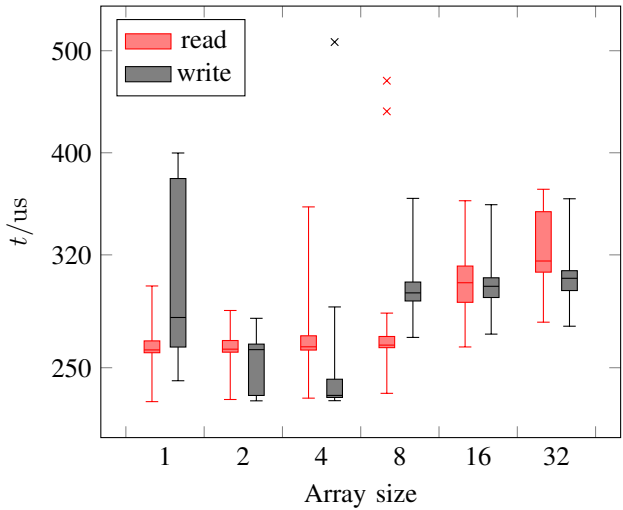
\begin{figure}
    \centering
    \resizebox{0.95\columnwidth}{!}{%
    \begin{tikzpicture}
        \begin{axis}[ymode=log,log basis y={10},
            ytick ={250000,320000,400000,500000,650000},
            yticklabels ={250,320,400,500,650},
            boxplot/draw direction=y,
            xlabel={Array size},
            xtick={0,1,2,...,50},
            x tick label as interval,
            xticklabels={1,2,4,8,16,32},
            ylabel={$t/\mathrm{us}$},
            legend pos={north west},
            cycle list={{red},{black}},
            log ticks with fixed point,
            area legend,
            boxplot={
                  %
                  %
                  draw position={1/3 + floor(\plotnumofactualtype/2) + 1/3*mod(\plotnumofactualtype,2)},
                  %
                  box extend=0.2,
              },
            every axis plot/.append style={fill,fill opacity=0.5},
        ]
        \addplot +[mark=x,boxplot prepared={lower whisker=232137.950000,lower quartile=258339.000000,median=259927.000000, upper quartile=265115.000000,upper whisker=298949.020000},] coordinates {
};
\addplot +[mark=x,boxplot prepared={lower whisker=242984.760000,lower quartile=261566.000000,median=279035.000000, upper quartile=378149.000000,upper whisker=399859.480000},] coordinates {
};
\addplot +[mark=x,boxplot prepared={lower whisker=233223.850000,lower quartile=258654.000000,median=260365.000000, upper quartile=265338.000000,upper whisker=283347.530000},] coordinates {
};
\addplot +[mark=x,boxplot prepared={lower whisker=232548.380000,lower quartile=235289.250000,median=260095.500000, upper quartile=263246.500000,upper whisker=278535.080000},] coordinates {
};
\addplot +[mark=x,boxplot prepared={lower whisker=233913.490000,lower quartile=259850.750000,median=261715.500000, upper quartile=268087.000000,upper whisker=355399.620000},] coordinates {
};
\addplot +[mark=x,boxplot prepared={lower whisker=232622.950000,lower quartile=234273.750000,median=235332.000000, upper quartile=243794.000000,upper whisker=285566.320000},] coordinates {
(0,509588)
};
\addplot +[mark=x,boxplot prepared={lower whisker=236397.930000,lower quartile=261217.250000,median=262675.000000, upper quartile=267688.750000,upper whisker=281738.170000},] coordinates {
(0,468288)
(0,437946)
};
\addplot +[mark=x,boxplot prepared={lower whisker=267136.480000,lower quartile=289300.250000,median=294476.000000, upper quartile=301531.000000,upper whisker=362096.580000},] coordinates {
};
\addplot +[mark=x,boxplot prepared={lower whisker=261627.560000,lower quartile=288453.250000,median=301066.500000, upper quartile=312257.000000,upper whisker=360194.440000},] coordinates {
};
\addplot +[mark=x,boxplot prepared={lower whisker=269063.040000,lower quartile=291486.000000,median=298726.500000, upper quartile=304339.250000,upper whisker=357089.340000},] coordinates {
};
\addplot +[mark=x,boxplot prepared={lower whisker=276192.590000,lower quartile=308128.500000,median=315739.500000, upper quartile=351674.000000,upper whisker=369368.410000},] coordinates {
};
\addlegendentry{read}
\addplot +[mark=x,boxplot prepared={lower whisker=273763.990000,lower quartile=295966.250000,median=304018.000000, upper quartile=309121.750000,upper whisker=361793.550000},] coordinates {
};
\addlegendentry{write}

       \end{axis}
    \end{tikzpicture}%
    }
    \caption{GRPC Client \SI{32}{\bit} access latency boxplot for small transfer sizes. The box represents the quartiles, the whiskers the \SI{99}{\percent}-quantile. (N=1000 averages)}
    \label{fig:client-singlecall-boxplot}
\end{figure}

As one can expect, the access time over network is significantly increased. Measurement results for the duration of register read and write operations via the gRPC client are given in \cref{fig:client-singlecall} and \cref{tab:register-latency}. Also the latency remains constant at first because only single packets are transmitted. For larger transfers the bandwidth is limited due to the hardware access latency (read operations with \SI{81}{\mega\bit\per\second}). If the packets are written to arrays in software the throughput is limited to \SI{314}{\mega\bit\per\second} for read access. The combined write and read access is handled within a single remote procedure call and is therefore similar to the other duration values. \cref{fig:client-singlecall-boxplot} shows the boxplot of network access for small data sizes. Since Ethernet itself is not a deterministic protocol\cite{7005074}, the spread of the distribution is widened.

\subsection{Data Transfer Speed}
For limited data forwarding during measurements, the gRPC interface can be used. The data can be transferred via a single gRPC call or with streams, that can be continuously read or written to. We investigate three different transfer methods, the single call transfer speed, a stream with \SI{32}{\bit} integer value arrays and a stream with plain byte arrays.

\begin{figure}
    \centering
    \resizebox{0.95\columnwidth}{!}{%
    \begin{tikzpicture}
        \begin{axis}[xmode=log,log basis x={2},
            xlabel={Blocksize in bytes},xtick={1024,2048,4096,8192,16384,32768,65536,131072},
            ylabel={Mbit/s},
            legend pos={north west},
        ]
        \addplot [color=blue, only marks,mark=o,]
                 plot [error bars/.cd, y dir = both, y explicit]
                 table[x index=0, y index=9,
                 col sep=comma]{grpc_32bitstream_read.csv};
        \addlegendentry{\SI{4}{B} read\,\,}
        \addplot [color=blue!50!yellow, only marks,mark=o,]
                 plot [error bars/.cd, y dir = both, y explicit]
                 table[x index=0, y index=9,
                 col sep=comma]{grpc_32bitstream_write.csv};
        \addlegendentry{\SI{4}{B} write}
        \addplot [color=red, only marks,mark=+,]
                 plot [error bars/.cd, y dir = both, y explicit]
                 table[x index=0, y index=9,
                 col sep=comma]{grpc_bytestream_read.csv};
        \addlegendentry{\SI{1}{B} read\,\,}
        \addplot [color=red!50!yellow, only marks,mark=+,]
                 plot [error bars/.cd, y dir = both, y explicit]
                 table[x index=0, y index=9,
                 col sep=comma]{grpc_bytestream_write.csv};
        \addlegendentry{\SI{1}{B} write}
       \end{axis}
    \end{tikzpicture}%
    }
    \caption{GRPC transfer spreed measurement. A preallocated and prefilled array of \SI{32}{\bit} and \SI{8}{bit} arrays were transfered between the ZynqMP platform and the client PC. Each block size was sent $2^{10}$ times through the stream. (N=10 averages)}
    \label{fig:transfer-speed}
\end{figure}
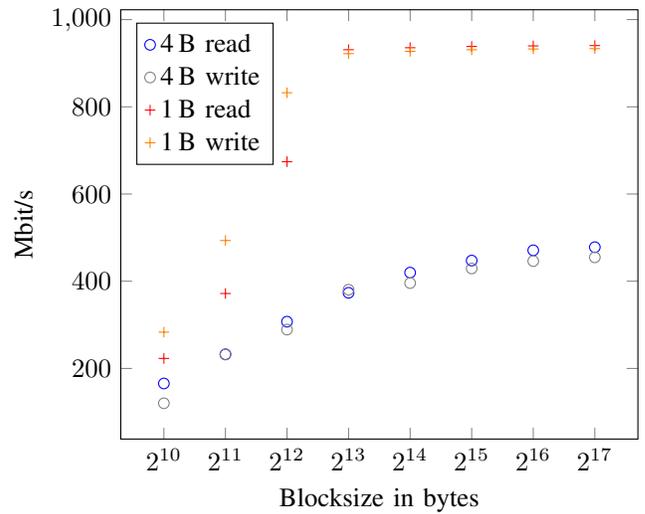

For flow rate measurements, all compression in the gRPC channel was disabled and transmitted arrays were filled with random data. This suppresses possible optimizations in serialization. 
The measured data throughput is shown in \cref{fig:transfer-speed}. The transfer with byte arrays results in speeds of 940 / \SI{934}{\mega\bit\per\second} (read/write)\cite{GAMESS2008585}. This is much faster than with \SI{32}{\bit} arrays, which are only transferred with 456 / \SI{479}{\mega\bit\per\second}. During the transfer, one of the cores of the ARM processor is fully utilized, the difference in transfer rates could be caused by the more demanding marshalling of \SI{4}{\byte} integers.

\subsection{I2C Latency}
Although i$^2$c is a bus system that is among the scope of slow control, its speed is an important parameter to benchmark. External devices such as control voltage generators monitoring devices utilize this bus and are relevant for system calibration. The measurement setup uses the on-die Cadence controller and an Analog Devices AD5675R digital-to-analog converter chip is used that is connected to the i$^2$c PMOD connector of the ZCU102 board. This connector is directly attached to the SDA/SCL pins, with a voltage level shifter but without any multiplexer in between. In order to benchmark the latency, the \texttt{ad5696-i2c} iio kernel driver is used in combination with a Sysfs\-Endpoint to write a \SI{16}{\bit} value to the DAC registers. 

\begin{table}[h]
    \centering
    \caption{Single \SI{16}{\bit} latencies of an i$^2$c device via the Linux Kernel driver and the sysfs interface. Comparison between direct endpoint access and access from a gRPC client via Ethernet (N=10000 averages). The client minimum is the smallest outlier of the distribution.}
    \begin{tabular}{c|c|c|c}
     & \textbf{Endpoint} & \textbf{Client} & \textbf{Client minimum} \\ 
    \hline
    Read(R) & \SI{257(2)}{\micro\second} & \SI{805(47)}{\micro\second} &  \SI{614}{\micro\second} \\ 
    Write(W) & \SI{161(1)}{\micro\second} & \SI{472(10)}{\micro\second} &  \SI{396}{\micro\second} \\ 
    W\&R & \SI{413(2)}{\micro\second} & \SI{909(64)}{\micro\second} & \SI{674}{\micro\second}
    \end{tabular}
    \label{tab:i2c-latency}
\end{table}

In \cref{tab:i2c-latency} the latency for accesses on the device and from client are shown. The register access time for i$^2$c is in the same order of magnitude as for Ethernet. The minimum latencies of distribution compare well to the single Ethernet and i$^2$c latency. For the larger part of transfers the combination shows an increase of client access latency.

\section{Conclusion}
\label{sec:conclusion}
We developed and presented a control software for Linux based SoC-FPGA systems called ServiceHub. It is specifically targeted on heterogeneous electronics systems that integrate different hardware components and need flexible means for external configuration and data access. The software is already actively used in measurement systems for the readout of superconducting sensors and quantum bits. The modular approach includes a gRPC server and allows the definition of plugins which can be used to control single hardware modules or groups via gRPC. It also provides an abstraction layer that simplifies the input and output to platform devices or Linux kernel drivers. The access latency over the abstraction layer is in the range of \SI{482}{\nano\second} for register reads and \SI{280}{\nano\second} writes and over the gRPC interface data rates of up to \SI{941}{\mega\bit\per\second} read and \SI{934}{\mega\bit\per\second} write are possible. 

The gRPC based interface makes it easy to connect other clients from different programming languages and enables a flexible hierarchical abstraction of functionality. The achieved latencies and speeds are fully sufficient to control the device, but the bare metal benchmark suggests that latency can be further reduced. Also the use of a real time patch-set for the Linux kernel could improve the spikes in read and write operations within the endpoints. For higher data rates in future experiments the throughput can be increased by integrating a second network interface and RDMA-based solutions like RoCE\cite{rocev2}. The gRPC interface, however, completely utilizes the bandwidth of the \SI{1}{\giga\bit\per\second} Ethernet.

\section*{Acknowledgment}
Nick Karcher acknowledges the support by the Doctoral School \emph{Karlsruhe School of Elementary and Astroparticle Physics: Science and Technology}. Richard Gebauer acknowledges support by the State Graduate Sponsorship Program (LGF) and the Helmholtz International Research School for Teratronics (HIRST).



%
%
%

\balance

\bibliographystyle{unsrt}
\bibliography{bibliography}

\end{document}